\documentclass[letterpaper]{article} % DO NOT CHANGE THIS
\usepackage{aaai25}  % DO NOT CHANGE THIS
\usepackage{times}  % DO NOT CHANGE THIS
\usepackage{helvet}  % DO NOT CHANGE THIS
\usepackage{courier}  % DO NOT CHANGE THIS
\usepackage[hyphens]{url}  % DO NOT CHANGE THIS
\usepackage{graphicx} % DO NOT CHANGE THIS
\usepackage{natbib}  % DO NOT CHANGE THIS AND DO NOT ADD ANY OPTIONS TO IT
\usepackage{caption} % DO NOT CHANGE THIS AND DO NOT ADD ANY OPTIONS TO IT
\usepackage{amsfonts}
\usepackage{amsmath}
\usepackage{xcolor}
\usepackage{booktabs}
\usepackage{multirow}
\usepackage{adjustbox}   % For adjusting table size
\usepackage{caption}     % For customizing captions if needed
\usepackage{algorithm}
\usepackage{algorithmic}

\usepackage{newfloat}
\usepackage{listings}
\DeclareCaptionStyle{ruled}{labelfont=normalfont,labelsep=colon,strut=off} % DO NOT CHANGE THIS
\floatstyle{ruled}
\newfloat{listing}{tb}{lst}{}
\floatname{listing}{Listing}
\title{Combating Phone Scams with LLM-based Detection: Where Do We Stand?}

\author{
    Zitong Shen,
    Kangzhong Wang,
    Youqian Zhang\thanks{Corresponding author: you-qian.zhang@polyu.edu.hk},
    Grace Ngai,
    Eugene Y. Fu
}
\affiliations{
    The Hong Kong Polytechnic University, Hung Hom, Kowloon, Hong Kong\\
    esther.shen@connect.polyu.hk
}
\begin{document}

\maketitle

\begin{abstract}
Phone scams pose a significant threat to individuals and communities, causing substantial financial losses and emotional distress. 
Despite ongoing efforts to combat these scams, scammers continue to adapt and refine their tactics, making it imperative to explore innovative countermeasures. 
This research explores the potential of large language models (LLMs) to provide detection of fraudulent phone calls. 
By analyzing the conversational dynamics between scammers and victims, LLM-based detectors can identify potential scams as they occur, offering immediate protection to users. 
While such approaches demonstrate promising results, we also acknowledge the challenges of biased datasets, relatively low recall, and hallucinations that must be addressed for further advancement in this field.
\end{abstract}

% Uncomment the following to link to your code, datasets, an extended version or similar.
%
% \begin{links}
%     \link{Code}{https://aaai.org/example/code}
%     \link{Datasets}{https://aaai.org/example/datasets}
%     \link{Extended version}{https://aaai.org/example/extended-version}
% \end{links}

\section{Introduction}
\label{sec:introduction}

Phone scams have become an increasingly pervasive threat in recent years, causing significant financial harm, as well as emotional distress to many people. 
Despite efforts by authorities and companies, such as improving public awareness, e.g., education~\cite{burke2022can}, and implementing anti-scam tools, e.g., robocall detection and blacklists~\cite{pandit2023combating}, scammers can still devise sophisticated tactics to evade these protection mechanisms, and continue to deceive the victims.
For example, scammers can use spoofed phone numbers to appear legitimate, and further employ a range of deceptive strategies, including psychological manipulation, fear-mongering, and the use of advanced technologies like deepfake audio and video, to force the victims into compliance. 
%A recent report highlighted the devastating impact of these scams, with one scammer leveraging artificial intelligence (AI) to pretend to be the boss from the headquarter, resulting in a staggering 25 million USD loss~\cite{chen2024finance}.

The rapid advancement of large language models (LLM) has ignited the hope that it can effectively detect phone scams by analyzing call dialogues. 
However, this seemingly straightforward approach is fraught with challenges. 
Beyond pricacy concerns, this paper delves deeper into the practical challenges of using LLM for phone scam detection, focusing on the following three aspects:

\begin{itemize}
 \item \textbf{Biased Datasets:} Models trained on biased datasets can be overly sensitive to specific keywords, making them vulnerable to adversarial tactics.
 \item \textbf{Low Recall:} A low recall rate means more false negatives, meaning that scammers have a high chance of evading detection. 
 \item \textbf{Hallucinations:} Hallucinations in LLMs can hinder accurate scam detection, especially in complex or ambiguous conversations. This can result in false positives or negatives, compromising the system's effectiveness.
\end{itemize}

% \begin{figure}[t]
% \centering
% \includegraphics[width=0.45\textwidth]{figures/methodology.png}
% \caption{Our proposed methods for fraudulent call detection}
% \label{fig:methodology}
% \end{figure}

%\begin{figure}[t]
%\centering
%\includegraphics[width=0.47\textwidth, height=0.13\textheight]{figures/methodology.png}
%\caption{Our proposed methods for fraudulent call detection}
%\label{fig:methodology}
%\end{figure}

\section{Analysis of Existing Datasets}

\subsection{Data Collection}
We obtained all available scam datasets from public sources~\cite {huggingface_bothbosu}, which contain dialogue transcripts between scammers and victims.
A summary of these datasets are shown in the supplementary document, and they are labeled as ``SC'', ``SD'', and ``MASC'', respectively.
Since these datasets are synthesized, we collected authentic fraudulent calls (i.e., ``Our-Real'') from available video recordings sourced from public platforms such as YouTube, and used speech-to-text technology to convert the speech from both the scammers and the victims into transcripts. 
In addition, we also generated our own synthesized dataset (i.e., ``Our-Synt''), and its details will be presented later.
\subsection{Bias in Datasets}

To identify distinctive features that could differentiate between scam and normal conversations, we conducted a comprehensive analysis of the datasets, excluding ``Our-Synt''. 
Utilizing Term Frequency-Inverse Document Frequency (TF-IDF) analysis, we extracted the top 50 most informative terms.
These extracted features were then used to train several classic machine learning classifiers, including Support Vector Machines (SVM), Random Forest (RF), and k-Nearest Neighbors (KNN). 
We only show the results of RF in Table~\ref{tab:ML_performance} as results across different models are similar.
All classifiers achieving exceptionally high performance metrics, consistently exceeding 0.99.

However, a deeper investigation revealed that this success was primarily attributed to the presence of specific terms that frequently appeared in fraudulent conversations. 
Scammers can easily circumvent these models by strategically avoiding these identified terms.
To validate this hypothesis, we generated new fraudulent and genuine conversations using Claude-3.5. 
These synthetic conversations (i.e., ``Our-Synt'') were carefully crafted to ensure that both fraudulent and genuine samples incorporated the top 50 extracted terms from real conversations.
When these newly generated samples were classified using the previously trained models, we observed a dramatic decline in performance, showing that scammers can exploit these vulnerabilities by adapting their tactics to avoid detection.

\begin{table}[t]
    \centering
    % \raggedright
    \caption{Random Forests' Performance on Datasets}
    \label{tab:ML_performance}
    \setlength{\tabcolsep}{2pt}
    \footnotesize 
    \begin{tabular}{llllll}
        \toprule
        & \textbf{SC} & \textbf{SD} & \textbf{MASC} & \textbf{Our-Real} & \textbf{Our-Synt} \\
        \midrule
        Accuracy  & 1.00 & 1.00 & 0.99 & 1.00 & 0.53 \\
        Precision    & 1.00 & 1.00 & 0.99 & 1.00 & 0.62 \\
        Recall    & 1.00 & 1.00 & 1.00 & 1.00 & 0.16 \\
        F1-Score   & 1.00 & 1.00 & 0.99 & 1.00 & 0.26 \\
        \bottomrule
    \end{tabular}
\end{table}

\section{LLM-based Detector}

\begin{figure}[t]
\centering
\includegraphics[width=0.43\textwidth]{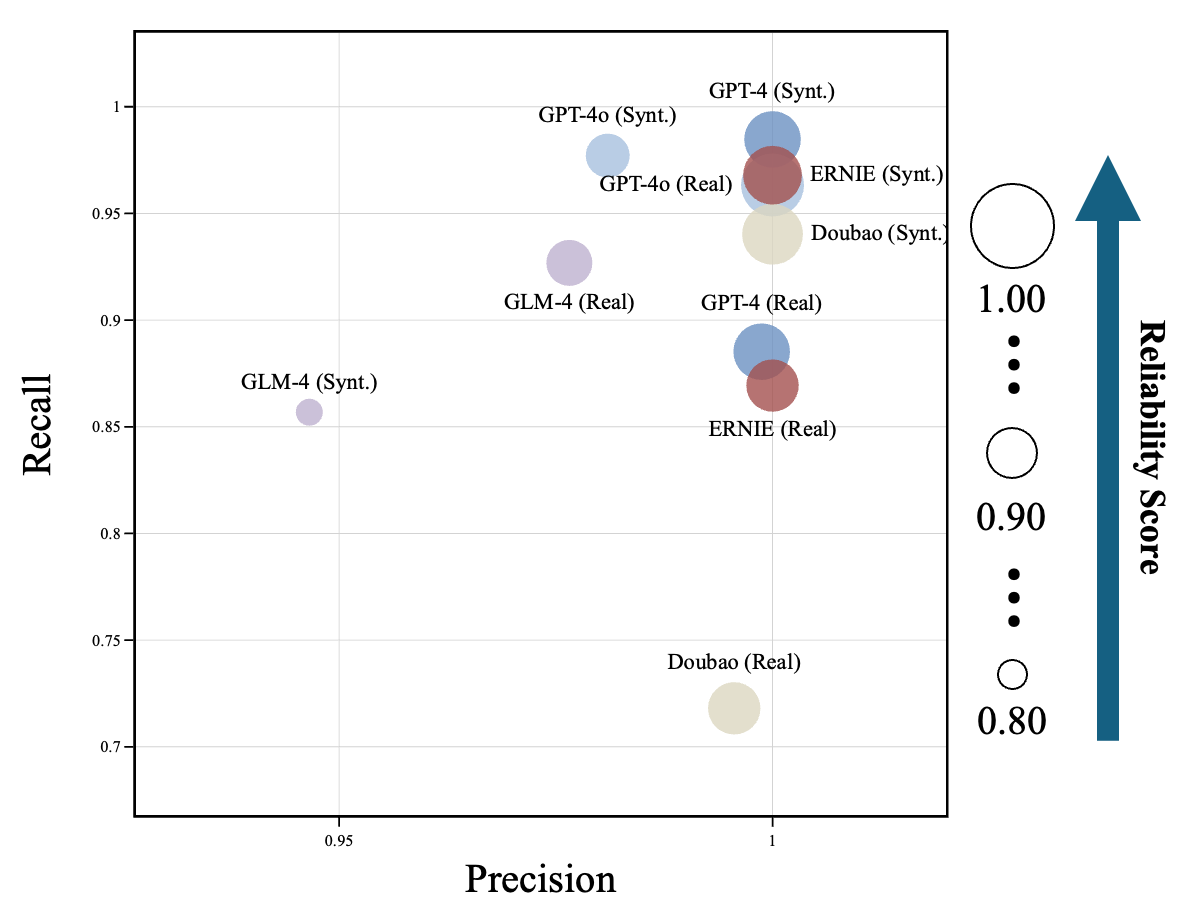}
\caption{Performance of different LLM models.}
\label{fig:RRP}
\end{figure}

Given LLMs' text understanding capabilities, we explored their potential to assist in identifying fraudulent conversations. 
We employed our datasets, and selected five popular LLMs: GPT-4, GPT-4o, GLM4, Doubao-Pro-32k, and ERNIE-3.5-8k. 
For each sample, we provided the prompt to LLMs: ``Please determine if this conversation exhibits any fraudulent tendencies. Provide a yes or no answer and explain your reasoning.''
The experimental process was repeated five times, and the averaged results are presented in Figure~\ref{fig:RRP}.

For ``Our-Real'', the LLMs achieved a precision of over 0.98 but the recall can be as low as 0.72. 
For synthetic data, precision exceeded 0.95, and recall can be as low as 0.86. 
Compared to traditional classifiers, LLMs demonstrated superior performance, as LLMs may not rely on keywords only, but can better understand the subtleties of language used by scammers, such as emotional manipulation or persuasive techniques.
%Also, LLMs can analyze the entire conversation, considering the context of previous exchanges to identify patterns and inconsistencies that may indicate fraudulent activity.

% \begin{table}[htbp]
%     \centering
%     % \caption{LLMs' Performances on Real and Synthetic Datasets}
%     \caption{Different LLMs on Real and Synthetic Datasets}
%     \label{tab:ML_performance_combined}
%     \setlength{\tabcolsep}{2pt} % 进一步减少列间的水平间距
%     \footnotesize % 仅减小表格内的字体大小
%     \begin{tabular}{ccccccc}
%         \toprule
%          & \textbf{Dataset} & \textbf{GPT4} & \textbf{GPT4o} & \textbf{GLM4} & \textbf{Doubao} & \textbf{ERNIE} \\
%         \midrule
%         \multirow{2}{*}{\textbf{Accuracy}}    & Real       & 0.94 & 0.98 & 0.96 & 0.86 & 0.93 \\
%                                      & Syn.       & 0.99 & 0.98 & 0.91 & 0.97 & 0.98 \\
%         \midrule
%         \multirow{2}{*}{\textbf{Precision}}   & Real       & 1.00 & 1.00 & 0.98 & 1.00 & 1.00 \\
%                                      & Syn.       & 1.00 & 0.98 & 0.95 & 1.00 & 1.00 \\
%         \midrule
%         \multirow{2}{*}{\textbf{Recall}}      & Real       & 0.89 & 0.96 & 0.93 & 0.72 & 0.87 \\
%                                      & Syn.       & 0.98 & 0.98 & 0.86 & 0.94 & 0.97 \\
%         \midrule
%         \multirow{2}{*}{\textbf{F1-Score}}          & Real       & 0.94 & 0.98 & 0.96 & 0.84 & 0.93 \\
%                                      & Syn.       & 0.99 & 0.98 & 0.91 & 0.97 & 0.98 \\
%         \bottomrule
%     \end{tabular}
% \end{table}

\subsection{Relatively Low Recall}
A low recall of 0.72 implies that 28 out of 100 fraudulent calls could still go undetected. 
For a task as critical as scam detection, a recall approaching 1 is desirable to minimize false negatives. 
However, this must be balanced with precision to avoid misclassifying genuine conversations.

\subsection{Hallucination}
Additionally, our experiments repeated for 5 times revealed inconsistencies in model responses.
The same sample might be classified differently in subsequent runs. 
To quantify this, we introduced a reliability score based on discrete semantic entropy~\cite{farquhar2024detecting}. 
We define the reliability score as:
\[
\text{Reliability~Score} = 1 - \frac{\sum_{i = 1}^{N}\sum_{j = 1}^{M}p_{i,j}\ln p_{i, j}}{N \times 0.5\ln 0.5}
\]
where $N$ is the number samples, and $M$ is the number of classes, and $p$ is the probability of a certain class.
The reliability score is normalized into the range between 0 and 1, and a higher value indicates greater model reliability, or in other words fewer hallucinations.

As illustrated in Figure~\ref{fig:RRP}, GPT-4o exhibited exceptional reliability on the ``Our-Real" dataset, achieving a perfect reliability score of 1. However, other models demonstrated varying levels of reliability. Notably, GLM-4 on the ``Our-Synt" dataset achieved a reliability score of only 0.83, indicating a higher propensity for inconsistent decisions on certain samples. These findings underscore the importance of carefully considering model reliability when deploying AI-based scam detection systems.

\section{Conclusion}
% Our investigation into LLM-powered phone scam detection revealed both its promises and limitations. 
% While traditional machine learning models achieved high initial performance on biased datasets, they were susceptible to adversarial attacks. 
% LLMs, with their advanced text understanding capabilities, demonstrated superior performance in identifying real-world scams. 
% However, challenges remain, including the trade-off between precision and recall, and the potential for LLM hallucinations, and future work are essential to get rid of these challenges.

Our investigation into LLM-powered phone scam detection places us at a critical juncture, balancing promises with persistent challenges. While traditional machine learning models achieved high initial performance on biased datasets, they were susceptible to adversarial attacks. 
LLMs, with their advanced text understanding capabilities, demonstrated superior performance in identifying real-world scams. 
However, challenges remain, including the trade-off between precision and recall, and the potential for LLM hallucinations, and future work are essential to get rid of these challenges.

\bibliography{aaai25}

\end{document}